\pdfoutput=1

\documentclass[11pt]{article}

\usepackage[final]{acl}

\usepackage{times}
\usepackage{latexsym}
\usepackage{multirow}
\usepackage[T1]{fontenc}

\usepackage[utf8]{inputenc}

\usepackage{microtype}
\usepackage{amssymb}
\usepackage{xcolor}
\usepackage{times}
\usepackage{soul}
\usepackage{url}
\usepackage{hyperref}
\usepackage{enumitem}
\usepackage{anyfontsize}
\usepackage[utf8]{inputenc}
\usepackage{booktabs}
\usepackage{amsmath}
\usepackage{tikz}
\usepackage{subcaption}

\usepackage{graphicx}
\usepackage{adjustbox}
\usepackage{algorithm}
\usepackage{algorithmic}

\title{LOGIC-LM++: Multi-Step Refinement for Symbolic Formulations}
\author{
 \textbf{Shashank Kirtania},
 \textbf{Priyanshu Gupta},
 \textbf{Arjun Radhakrishna}
\\ \textbf{Microsoft} \\
\{t-skirtania, priyansgupta, arradha\} @microsoft.com
}

\newcommand{\logiclmpp}{Logic-LM++}
\newcommand{\logiclm}{Logic-LM}
\newcommand{\folio}{FOLIO}

\begin{document}
\maketitle

\begin{abstract}
In this paper we examine the limitations of Large Language Models (LLMs) for complex reasoning tasks. Although recent works have started to employ formal languages as an intermediate representation for reasoning tasks, they often face challenges in accurately generating and refining these formal specifications to ensure correctness. To address these issues, this paper proposes \logiclmpp{}, an improvement on \logiclm{} \cite{logic-lm}. It uses the ability of LLMs to do pairwise comparisons, allowing the evaluation of the refinements suggested by the LLM. The paper demonstrates that \logiclmpp{} outperforms \logiclm{} and other contemporary techniques across natural language reasoning tasks on three datasets, \folio{}, ProofWriter and AR-LSAT, with an average improvement of 18.5\% on standard prompting, 12.3\% on chain of thought prompting and  5\% on \logiclm{}.
\end{abstract}
\section{Introduction}
\label{sec:Intro}
Large language models (LLMs) have shown proven capability of reasoning \cite{DBLP:GPT3, chowdhery2022palm} but still struggle at complex reasoning problems as seen in real world assessments \cite{DBLP:AR-LSAT}. For complex multi hop reasoning tasks current state of the art approaches \cite{logic-lm, satlm} leverage formal languages as intermediate representation of these reasoning problems and utilize symbolic reasoners to come up with the right response. A typical workflow of such techniques consist of 3 steps: a natural language prompt which consist of the task information, a response formulation for the problem, final response generated with symbolic executor. \\

\begin{figure}[h!]
  \centering
  \begin{minipage}{0.5\textwidth}
    \centering
    \includegraphics[width=\columnwidth]{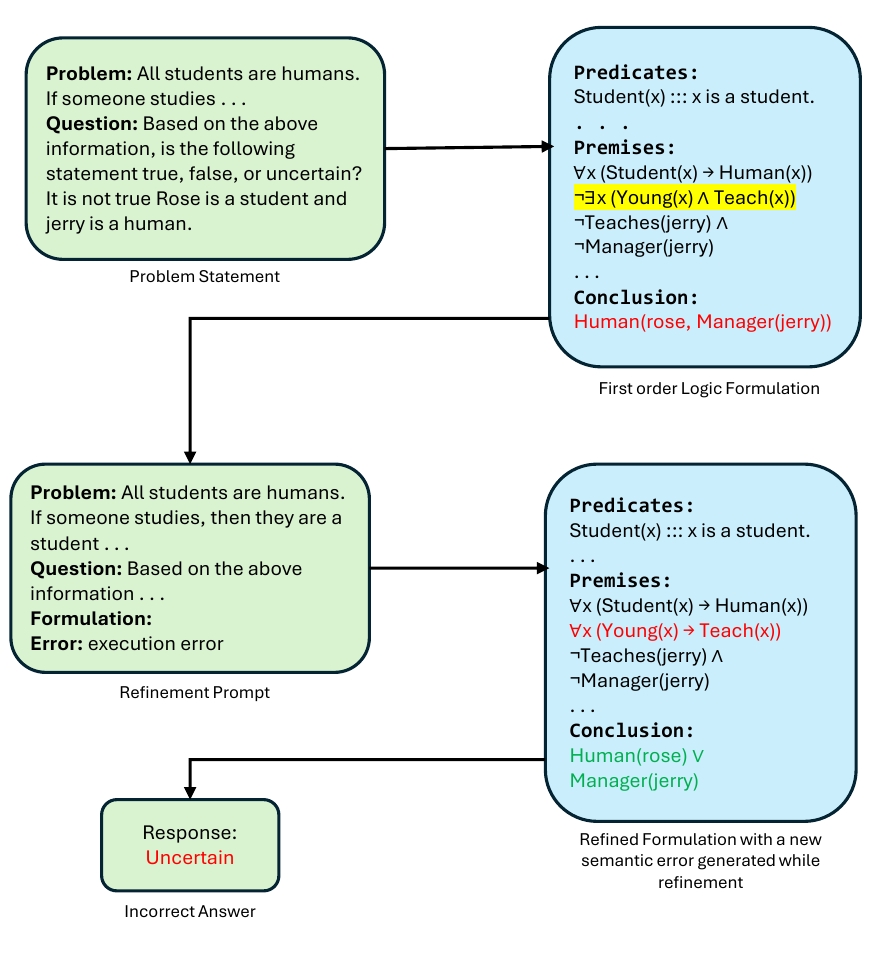}
    \caption{Refinement of logical formulations in \logiclm{}}
    \label{fig:failing-case}
  \end{minipage}\hfill
  \end{figure}

While logic-assisted LLM reasoning techniques are promising, we observe following problems in such systems: Firstly, LLMs are poor at generating intermediate formal specifications. A few techniques try to counter this problem with a refinement loop \cite{self-refine, selfcheck, reflexion} to improve upon the syntactical correctness of the symbolic formulation. Secondly, the LLMs are poor at repairing the formal representations with limited information with error information. For example, in Figure \ref{fig:failing-case} the LLM initially generates a syntactically incorrect formulation. After a turn of refinement, while the LLM is able to generate a response that is syntactically correct, it introduces a \emph{semantic} error in the formulation by incorrectly translating the statement "No young person teaches". These kind of incorrect translations from Natural Language (NL) to intermediate formal specifications is a common problem we observe over the failing cases of refinement. Thirdly, we observe that refinements are not always linear-resolving an error with the symbolic formulation can take multiple steps of careful edits and evaluation. The formulations generated in refinement stage in \ref{fig:failing-case} introduced the wrong interpretation of "No young person teaches" to "All young people teaches". \\
To address these challenges we propose to add following measures in \logiclm{} to enhance it's capabilities resulting in improved variant \logiclmpp{}.

We leverage the ability of LLMs to do pairwise comparison \cite{llmasjudge}, this gives us an opportunity to evaluate the refinements suggested by the LLM and do a semantic check with respect to the problem statement to ensure if the edits in the symbolic formulation generated while refinement improve the formulation semantically not just syntactically.

We also improve on the refinement mechanism present in \logiclm{} to give more context of the problem statement during refinement stage, this eliminates cases where recommended edits are appalling and do not improve the formulation significantly.
\section{Related Work}
\label{sec:Related Work}
\subsection{Reasoning with LLMs}
Large Language model (LLM) - based reasoning techniques commonly entail the deconstruction of complex questions into a sequence of intermediate steps, often referred to as chains, before reaching the ultimate solution. This technique is a reflection of methods such as Chain of Thought (CoT) prompting and its variations, as shown in various studies \cite{CoT, llm-as-0-shot}. These methodologies require the meticulous segmentation of a problem into a chain of smaller, manageable chunks. Each of these chunks represents a step in the reasoning process, guiding the model towards a comprehensive solution.
The concept of the reflection loop, as explored in previous research \cite{reflexion, selfrefine}, offers a means of refining the reasoning by identifying and eliminating any flaws that may be introduced by the LLM during a reasoning step. This process enhances the inherent capability of the LLM to self-correct, contributing to more accurate and reliable outcomes.
\begin{figure}[h!]
  \begin{minipage}{0.5\textwidth}
    \centering
    \includegraphics[width=\columnwidth]{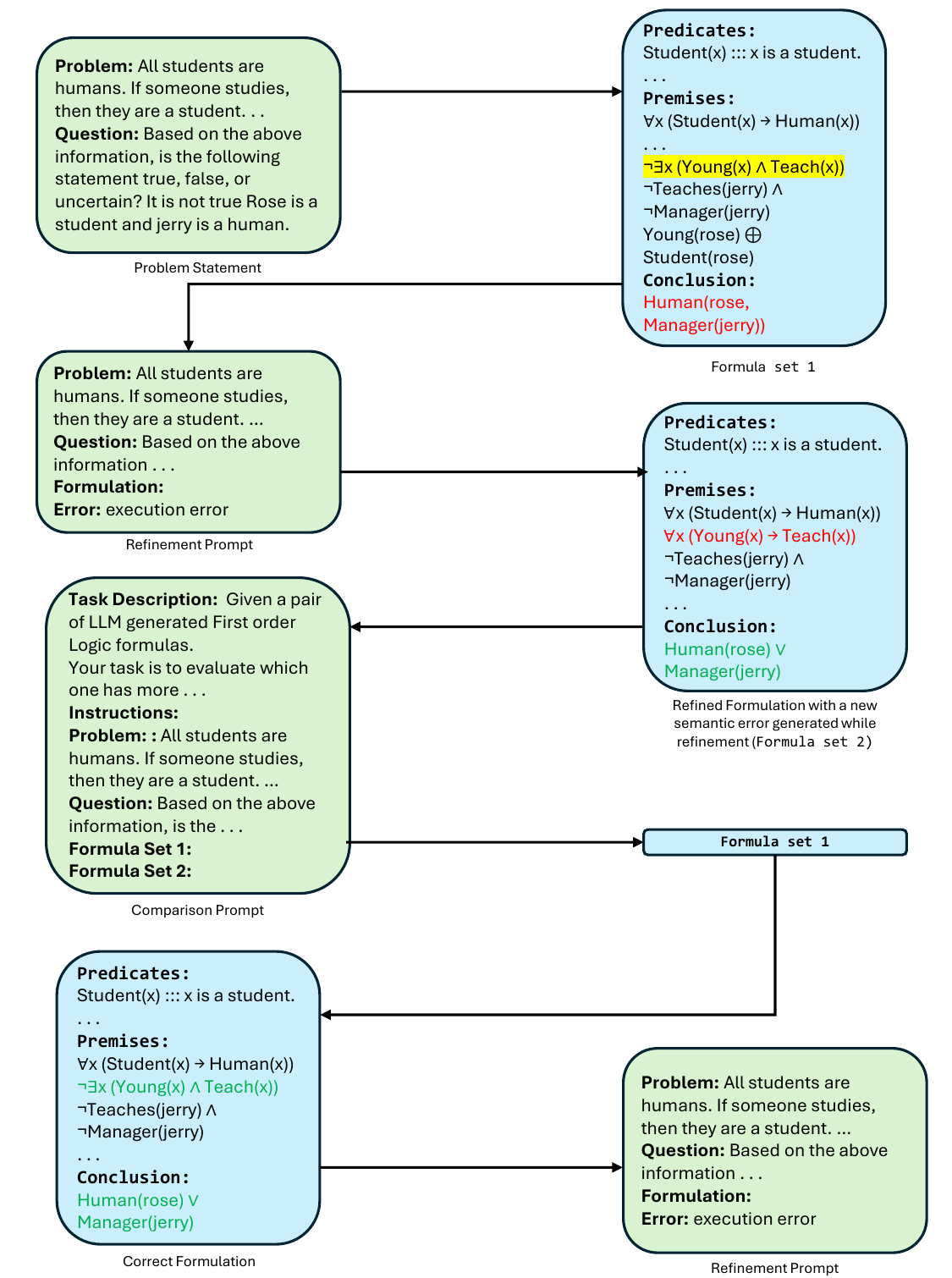}
    \caption{Improvement in refinement by \logiclmpp{}}
    \label{fig:winning-case}
  \end{minipage}
\end{figure}
Recent works have further explore the process of self-evaluation at these intermediate steps \cite{selfcheck, refiner}. This process involves the LLM assessing its reasoning at each step, allowing it to identify any inaccuracies. By rectifying these issues before proceeding to the next step, the LLM can ensure a more accurate and reliable chain of reasoning.
\begin{table*}[]
\centering
\resizebox{\textwidth}{!}{%
\begin{tabular}{@{}c|cccccccc@{}}
\toprule
\multirow{2}{*}{Dataset} & \multicolumn{4}{c}{\textbf{GPT-3.5 Turbo}} & \multicolumn{4}{c}{\textbf{GPT-4}} \\ \cmidrule(l){2-9} 
 & \textbf{Standard} & \textbf{CoT} & \textbf{\logiclm} & \multicolumn{1}{c|}{\textbf{\logiclmpp}} & \textbf{Standard} & \textbf{CoT} & \textbf{\logiclm} & \textbf{\logiclmpp} \\ \midrule
\folio & 45.09 & 57.35 & \textbf{62.80} & \multicolumn{1}{c|}{\textit{62.25}} & 69.11 & 70.58 & 78.92 & \textbf{84.80} \\
\multicolumn{1}{l|}{ProofWriter} & 35.50 & 49.17 & 58.33 & \multicolumn{1}{c|}{\textbf{58.83}} & 52.67 & 68.11 & 79.66 & 79.66 \\
AR-LSAT & 20.34 & 17.31 & 26.41 & \multicolumn{1}{c|}{\textbf{28.13}} & 33.33 & 35.06 & 43.04 & \textbf{46.32} \\ \bottomrule
\end{tabular}%
}
\caption{Accuracy of standard promoting, chain-of-thought (CoT) promoting, \logiclm{}  and \logiclmpp{}.}
\label{tab:full_result}
\end{table*}
Aligned with our objective of capturing the natural language intent of the user from symbolic formulations, recent works \cite{endres2024can} have also explored the translation of natural language into formal language post conditions. This research investigates how effectively we can convert the often-ambiguous language of human conversation into the precise, unambiguous language of formal logic. This translation process is crucial for the accurate interpretation and execution of user intent, particularly in complex or technical tasks.
\subsection{Tool-augmented Large Language Models}
Language models face inherent limitations, unable to access real-time data, execute actions, or conduct precise mathematical reasoning.To address this, recent research endeavors have sought to augment language models by integrating external resources such as retrievers \cite{DBLP:WebGPT,DBLP:REPLUG,DBLP:journals/corr/abs-2203-05115}, calculators \cite{DBLP:GSM8K}, code interpreters \cite{DBLP:Code4Struct}, planners \cite{DBLP:journals/corr/abs-2304-11477}, symbolic solvers \cite{satlm, logic-lm}, and other pretrained models \cite{DBLP:hugginggpt}. Notably, in the realm of mathematical reasoning, numerous investigations have illustrated the efficacy of incorporating calculators \cite{DBLP:GSM8K,DBLP:math_prompter} or Python interpreters \cite{DBLP:PAL,DBLP:POT} into language models, significantly enhancing performance by offloading numerical computations. Recent studies \cite{DBLP:PAL,DBLP:POT} have showcased improved effectiveness in arithmetic reasoning tasks by generating Python programs that delineate the reasoning process through sequenced chained commands.

\section{Methodology}
\label{sec: Methodology}
\subsection{Background}
\logiclm{} \cite{logic-lm} is a framework to decompose a reasoning problem into three stages:
\begin{enumerate}
    \item \emph{Problem Formulation}, where given a task description and a problem statement LLM write symbolic formulations that represents the NL problem. In Figure \ref{fig:failing-case} the NL prompt with task description is the problem formulator in \logiclm{}.
    \item\emph{Symbolic Reasoning}, where we use a symbolic solver like Prover92 \cite{Robinson1965}and Z3 theorem prover \cite{moura2008z3} to solve the formulations generated earlier.
    \item  \emph{Result interpretation}, where the produced output is mapped to the right answer using regex parsing.\end{enumerate}

\logiclm{} uses a refinement loop to fix errors in symbolic formulation at formulation and reasoning stages. However, \logiclm{} still struggles to improve on logical representations, showing almost no improvement after multiple iterations. Authors attribute this to semantic limitations of the formulation. To this end,
\logiclmpp{} aims to mitigate this limitation by improving the Logic-LM  refinement loop.
\subsection{Self-Refinement Agent}
Logic-LM defines the notion of a \emph{Self-Refinement Agent} to implement the refinement loop in the symbolic formulations in cases where the formulations did not yield a successful execution within the system. This agent is characterized by a \emph{refinement prompt} \ref{fig:failing-case}. In the original work, the refinement prompt constituted various few shot examples to act as exemplar for the model. While similar techniques have proven useful \cite{self-refine, reflexion}, we anecdotally observe that instead of helping the model it adds extra irrelevant information that distracts the model from fixing the issues relevant to the current formulation, consistent with similar studies in other domains \cite{logic-lm}. To alleviate this, instead of adding few-shots, we add the problem statement and the question to the refinement prompt alongside instructions to self-reflect on the model's failure to generate the right response. As we show later in Section \ref{sec: Results and Analysis}, this structure helps better \emph{contextualize} \cite{reflexion} the formulation to the self-reflection agent and help the system generate better refinements. 

\subsection{Backtracking Agent}
LLMs has shown remarkable results in automated evaluation benchmarks \cite{zheng2023judging} and has shown high alignment with the human judgement \cite{wei2024rethinking}. We use this capability of LLMs to assess if the repaired formulation by self-refinement improves the alignment of the human intent with LLM generated formulations. This allows us to get rid of the updates that are not helpful in future iterations and only use those updates where the changes help the model to come to the right formulation. 
\begin{figure*}
    \centering
  \includegraphics[width=\textwidth, height=0.3\textwidth]{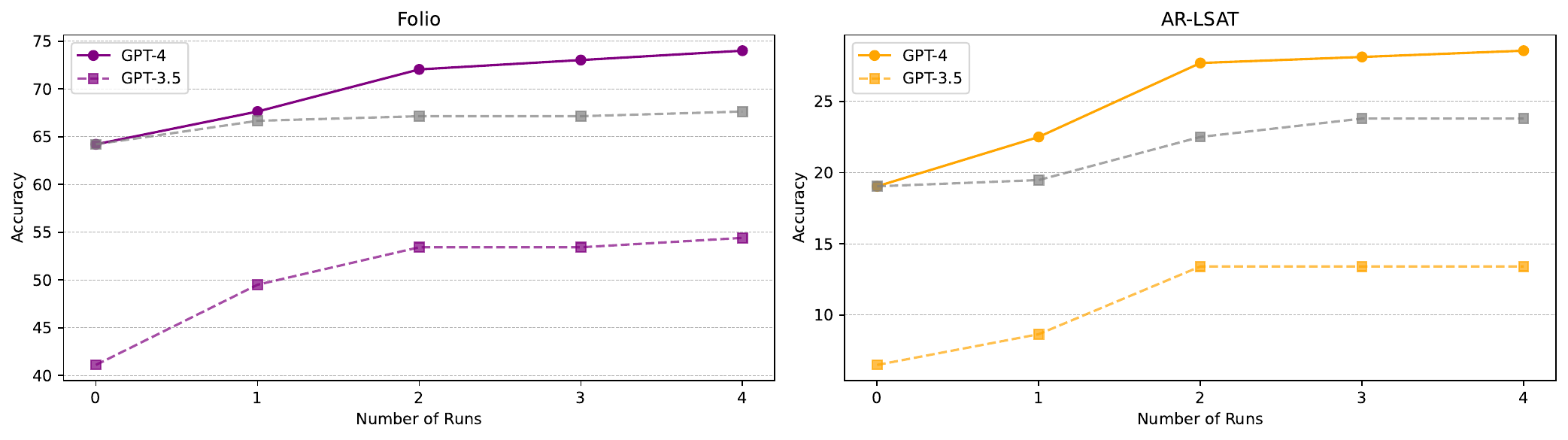}
  \caption{Accuracy in subsequent rounds of refinement. The grey line here represents the accuracy scores on self-refinement without backtracking with GPT-4.}
  \label{fig:acc_ineach_round}
\end{figure*}
In Figure \ref{fig:failing-case} we can see without the backtracking agent the LLM accepts the semantically incorrect symbolic formulations as the statement "No young person teaches" is translated to "all young people teach" since the code is syntactically correct there is no proof-check on the refinement.

However, In Figure \ref{fig:winning-case} we demonstrate in the same example with the backtracking agent \logiclmpp{} is able to generate right formulation by using the right formulation to represent "No young person teaches" and use the right formulation to describe the question "Rose is a student and Jerry is a human". This showcase how backtracking agent works as funnel to reduce the semantic error that is propagated at the refinement stage. In Figure \ref{fig:winning-case} we show on comparison of the two sets of the formulation, returns a more semantically correct formulation this allows \logiclmpp{} to only accept the edits if it improves or preserve the logical structure of the formulation.

\section{Experiments and Analysis}
\label{sec: Results and Analysis}
\subsection{Dataset}
\textbf{\folio{}}~\cite{DBLP:FOLIO} is a challenging expert-written dataset for logical reasoning. The problems are aligned with real-world knowledge and use natural wordings, and the questions require complex first-order logic reasoning to solve. We use the \folio{} test set for evaluation with 204 examples. 

\textbf{AR-LSAT}~\cite{zhong-etal-2022-analytical} is a dataset that collects all analytical logic reasoning questions from the Law School Admission Test from 1991 to 2016. We use the test set which has 231 multiple-choice questions. AR-LSAT is particularly challenging, with state-of-the-art model's performance slightly better than random guessing~\cite{DBLP:journals/corr/abs-2211-09110,DBLP:conf/iclr/Ribeiro0MZDKBRH23}. 

\textbf{ProofWriter} \cite{proofwriter} is another popular dataset on deductive logical reasoning. Since the problems are in more natual language like setting it makes semantic evaluation very relevant in the problem set. \logiclm{} use the open-world assumption (OWA) subset in which each example is a (problem, goal) pair and the label is one of {PROVED, DISPROVED, UNKNOWN}. \logiclm{} evaluate the pipeline with the hardest section of ProofWriter which contain total of 600 randomly sampled five step multi-hop reasoning questions.
\subsection{Principal Findings}
\begin{table}[]
\resizebox{\columnwidth}{!}{%
\begin{tabular}{@{}c|ccccc@{}}
\toprule
 &  & \multicolumn{2}{c}{\textbf{GPT-3.5 turbo}} & \multicolumn{2}{c}{\textbf{GPT-4}} \\ \midrule
\textbf{BT} &  & - & + & - & + \\ \midrule
\multirow{2}{*}{\textbf{FOLIO}} & $E_{\text{r}}$ & 84.3 & 84.3 & 85.8 & \textbf{86.7} \\
 & $E_{\text{a}}$ & 64.3 & \textbf{66.2} & 79.9 & 85.8 \\ \midrule
\multirow{2}{*}{\textbf{ProofWriter}} & $E_{\text{r}}$ & 95.6 & 95.6 & 99.0 & 99.0 \\
 & $E_{\text{a}}$ & 74.1 & \textbf{77.2} & \underline{79.6} & \underline{79.6} \\ \midrule
\multirow{2}{*}{\textbf{AR-LSAT}} & $E_{\text{r}}$ & 21.8 & \textbf{22.9} & \textbf{32.6} & 32.0 \\
 & $E_{\text{a}}$ & 60.3 & \textbf{64.1} & 60.0 & \textbf{66.2} \\ \bottomrule
\end{tabular}%
}
\caption{Execution rate ($E_{\text{r}}$) and Execution Accuracy ($E_{\text{a}}$) agent with Backtracking (BT).}
\label{tab:2}
\end{table}
We report the final results of \logiclmpp{} in Table \ref{tab:full_result}.
We try to answer 2 major research questions. \\
\textbf{RQ1: Can LLMs conduct pairwise comparisons of symbolic formulations based on their relevance to a natural language task description?}
LLMs have demonstrated promising capabilities in pairwise comparisons for NLG evaluations \cite{prometheus}, even in low-resource languages where their natural language generation abilities remain underdeveloped \cite{llmasjudge}. As depicted in Table \ref{tab:2}, the execution accuracy of the framework employing a backtracking agent is enhanced by approximately 6\% with GPT-4 and around 3\% with GPT3.5-turbo. Despite the average gain in execution rate being less than 1\%, these statistics underscore the empirical improvements in code quality in terms of semantic correctness.
Figure \ref{fig:failing-case} provides a working example from the \folio{} dataset. Although the code is syntactically correct after refinement, it misinterprets a logical statement. However, by implementing pairwise comparisons, the LLM can select the semantically correct formulation. This leads to the correct answer in the subsequent refinement iteration.

\textbf{RQ2: Does refinement by LLM always positively affect the formulations?}

In Figure \ref{fig:acc_ineach_round}, we evaluate the refinement process with and without backtracking. \logiclm{}'s accuracy plateaus with more runs because refined solutions may not represent the intended code. The author's also discuss this as a known limitation of the refinement process in the refinement loop they proposed. Backtracking, which reverts to the initial code if no semantic improvement is found, allows \logiclmpp{} to perform consistently better by continually reassessing and correcting refinements for more reliable results. 

\begin{figure}
  \centering
  \includegraphics[width=0.5\textwidth]{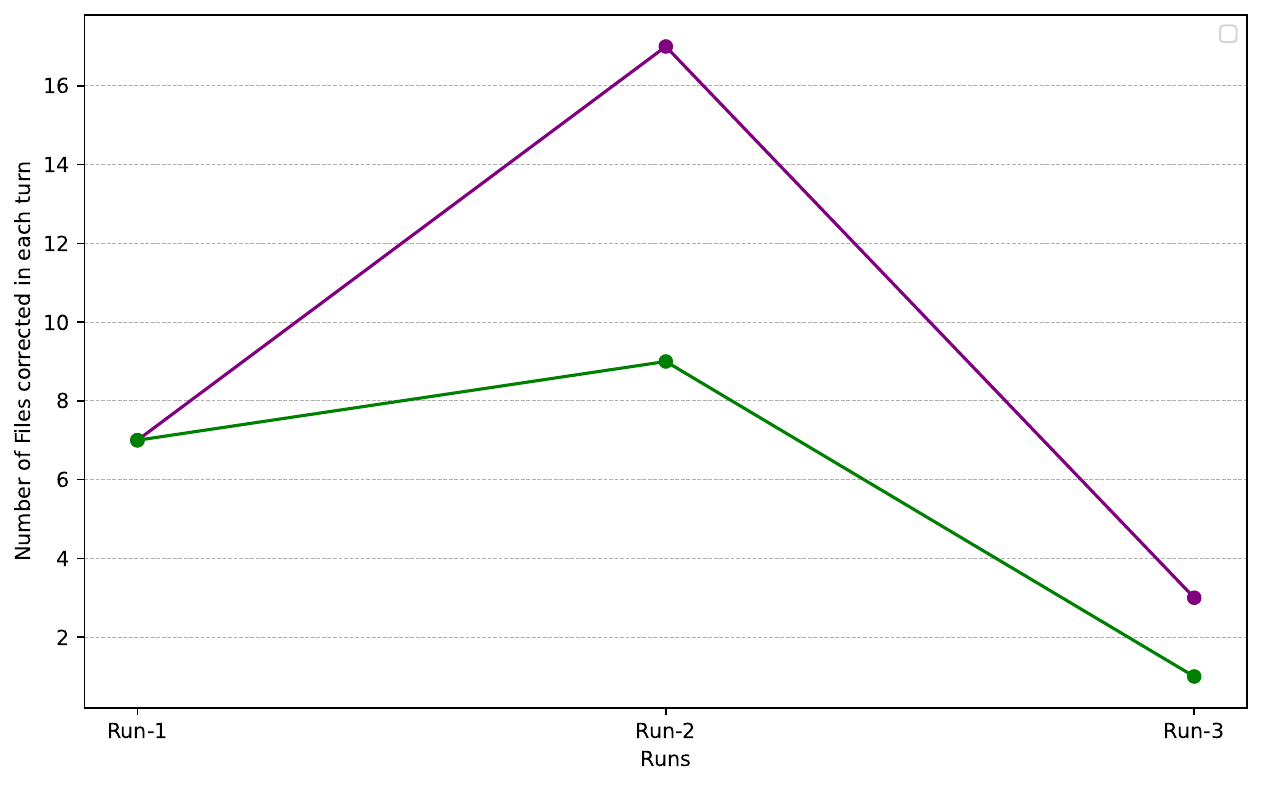}
  \caption{Number of symbolic formulations corrected after each turn of self-refinement with backtracking agent (purple) and without backtracking agent (green) in \folio{} with GPT-4.}
  \label{fig:fixed-across}
\end{figure}

Figure \ref{fig:fixed-across} shows that the backtracking agent significantly improves results in the second round within the \folio{} dataset, with a similar impact in later rounds. This indicates that backtracking is most effective early on since the generated refinement can also degrade the performance of the formulations, enabling \logiclmpp{} to achieve substantial better and iterative improvements over time.

\subsection{Error Analysis}
Even though \logiclmpp{} shows impressive improvements over standard refinement techniques, it still lacks behind in the cases where the first set of formulation generated is completely different from the  ground truth formulation. On analyzing the failure cases in \logiclm{} we note that the current pipeline relies a lot on fixing the bugs with current formulation without losing on semantic understanding, however in cases where the generating semantically correct formulations is hard the technique is contingent to initial formulations generated. 
\section{Discussion and Future Work}
\label{sec: Discussion}
Figure 3 reveals a significant observation regarding the iteration increase of Logic-LM, which appears to reach convergence substantially earlier than Logic-LM++. Logic-LM associates attributes this to the hard limit of semantically correctness that can be achieved with \logiclm{}.
The findings stress the importance of semantic accuracy, as the \logiclmpp{} exhibits consistently improved outcomes over multiple iterations, contrary to findings by \logiclm{}. This outcome is primarily attributed to the model's capability to revert to the initial formulation if the refined version does not offer a semantically superior representation.
Eventhough, \logiclmpp{} show promising results it only focus on symbolic formulations, this effort can be well generalised to other tool augmented techniques that rely on intermediate code representation with semantic improvements during refinement. 
\section{Conclusion }
\label{sec: Conclusion}
We propose \logiclmpp{} which beats state of the art results on natural language reasoning tasks on three datasets. \logiclmpp{} takes leverage of LLMs' reasoning capabilities to show significant improvements in efficient use of logic solvers for reasoning, we demonstrate that LLMs show promising results at conducting comparison between symbolic formulations even in cases where generating symbolic formulations is a hard task for LLM.
\section*{Limitation}
\label{sec: Limitation}
At present, \logiclmpp{} faces constraints in its capacity to effectively capture the semantic intricacies in reasoning tasks. This limitation notably complicates the evaluation process, especially when dealing with smaller LLMs like \cite{codellama}. The understanding required for accurate reasoning poses a significant challenge, particularly in contexts where the model's semantic comprehension may be insufficient. Due to this the assessment of performance becomes notably more complex. This limitation underscores the need for continued advancements in semantic understanding within LLMs to enhance their efficacy across reasoning tasks.
\bibliography{custom}
\end{document}